\documentstyle[12pt]{article}

\def\doit#1#2{\ifcase#1\or#2\fi}

\expandafter\ifx\csname amsppt.sty\endcsname\endinput
  \expandafter\def\csname amsppt.sty\endcsname{2.2 (2001/08/07)}\fi

\catcode`@=11
\catcode`@=12

\def\a{\alpha}

\def\S{\Sigma}  

\def\pmb#1{\setbox0=\hbox{${#1}$}%
   \kern-.025em\copy0\kern-\wd0
   \kern-.035em\copy0\kern-\wd0
   \kern.05em\copy0\kern-\wd0
   \kern-.035em\copy0\kern-\wd0
   \kern-.025em\box0 }

\def\bo{{\raise-.46ex\hbox{\large$\Box$}}} 

\def\TH{{\raise.2ex\hbox{$\displaystyle \bigodot$}\mskip-4.7mu %
\llap H \;}}
\def\face{{\raise.2ex\hbox{$\displaystyle \bigodot$}\mskip-2.2mu %
\llap {$\ddot
        \smile$}}}                           

\def\sp#1{{}^{#1}}                 

\def\Hat#1{\widehat{#1}}                        
\def\leftrightarrowfill{$\mathsurround=0pt \mathord\leftarrow
 \mkern-6mu
        \cleaders\hbox{$\mkern-2mu \mathord- \mkern-2mu$}\hfill
        \mkern-6mu \mathord\rightarrow$}
\def\dvec#1{\vbox{\ialign{##\crcr
        \leftrightarrowfill\crcr\noalign{\kern-1pt\nointerlineskip}
        $\hfil\displaystyle{#1}\hfil$\crcr}}}           
\def\dt#1{{\buildrel {\hbox{\LARGE .}} \over {#1}}}

\def\frac#1#2{{\textstyle{#1\over\vphantom2\smash{\raise.20ex
        \hbox{$\scriptstyle{#2}$}}}}}   
\def\sfrac#1#2{{\vphantom1\smash{\lower.5ex\hbox{\small$#1$}}\over
        \vphantom1\smash{\raise.4ex\hbox{\small$#2$}}}}
\def\bfrac#1#2{{\vphantom1\smash{\lower.5ex\hbox{$#1$}}\over
        \vphantom1\smash{\raise.3ex\hbox{$#2$}}}}       
\def\afrac#1#2{{\vphantom1\smash{\lower.5ex\hbox{$#1$}}\over#2}} 
\def\on#1#2{\mathop{\null#2}\limits^{#1}}       

\newskip\humongous \humongous=0pt plus 1000pt minus 1000pt

\newif\ifdtup

\doit0{
\def\ref#1{$\sp{#1)}$}
}

\parskip=\medskipamount          
\def\baselinestretch{1.2}       
\thicklines                         

\def\endtitle{\end{quotation}\newpage}  

\def\sect#1{\bigskip\medskip \goodbreak \noindent{\bf {#1}} %
\nobreak \medskip}
\def\refs{\sect{References} \footnotesize \frenchspacing \parskip=0pt}
\def\Item{\par\hang\textindent}

\def\[{\lfloor{\hskip 0.35pt}\!\!\!\lceil}
\def\]{\rfloor{\hskip 0.35pt}\!\!\!\rceil}

\def\plpl{{+\!\!\!\!\!{\hskip 0.009in}%
{\raise-1.0pt\hbox{$_+$}}  {\hskip 0.0008in}}}
\def\mimi{{-\!\!\!\!\!{\hskip 0.009in}%
{\raise-1.0pt\hbox{$_-$}}  {\hskip 0.0008in}}}

\def\hepth#1{\texttts{hep-th/{#1}}}

\def\texttts#1{\texttt{#1}}

\def\<<{<\!\!<} \def\>>{>\!\!>}
\def\Check#1{{\raise-1.0pt\hbox{\LARGE\v{}}{\hskip -10pt}{#1}}}

\def\eqques{{~\,={\hskip -11.5pt}\raise -1.8pt\hbox{\large ?}
{\hskip 4.5pt}}{}}

\def\fracmm#1#2{\,{{#1}\over{#2}}\,}

\def\frac#1#2{{\textstyle{#1\over\vphantom2\smash{\raise -.20ex
        \hbox{$\scriptstyle{#2}$}}}}}   

\def\scst{\scriptstyle}

\def\.{.$\,$}
\def\-{{\hskip 1.5pt}\hbox{-}}

\def\footnotes#1{{\hskip 1pt}\footnotemark$^)$\footnotetext{\hsize=6.5in $^)$~{#1}}}

\def\low#1{\hskip0.01in{\raise -3pt\hbox{${\hskip 1.0pt}\!_{#1}$}}}
\def\low#1{\hskip0.01in{\raise -3pt\hbox{$\!\!\!_{#1}$}}}
\def\ip{{=\!\!\! \mid}}

\begin{document}

\font\tenmib=cmmib10
\font\sevenmib=cmmib10 at 7pt 
\font\fivemib=cmmib10 at 5pt  
\font\tenbsy=cmbsy10
\font\sevenbsy=cmbsy10 at 7pt 
\font\fivebsy=cmbsy10 at 5pt  
\def\BMfont{\textfont0\tenbf \scriptfont0\sevenbf
                              \scriptscriptfont0\fivebf
            \textfont1\tenmib \scriptfont1\sevenmib
                               \scriptscriptfont1\fivemib
            \textfont2\tenbsy \scriptfont2\sevenbsy
                               \scriptscriptfont2\fivebsy}
\def\rlx{\relax\leavevmode}
\def\BM#1{\rlx\ifmmode\mathchoice
                      {\hbox{$\BMfont#1$}}
                      {\hbox{$\BMfont#1$}}
                      {\hbox{$\scriptstyle\BMfont#1$}}
                      {\hbox{$\scriptscriptstyle\BMfont#1$}}
                 \else{$\BMfont#1$}\fi}

\font\tenmib=cmmib10
\font\sevenmib=cmmib10 at 7pt 
\font\fivemib=cmmib10 at 5pt  
\font\tenbsy=cmbsy10
\font\sevenbsy=cmbsy10 at 7pt 
\font\fivebsy=cmbsy10 at 5pt  
\def\BMfont{\textfont0\tenbf \scriptfont0\sevenbf
                              \scriptscriptfont0\fivebf
            \textfont1\tenmib \scriptfont1\sevenmib
                               \scriptscriptfont1\fivemib
            \textfont2\tenbsy \scriptfont2\sevenbsy
                               \scriptscriptfont2\fivebsy}
\def\BM#1{\rlx\ifmmode\mathchoice
                      {\hbox{$\BMfont#1$}}
                      {\hbox{$\BMfont#1$}}
                      {\hbox{$\scriptstyle\BMfont#1$}}
                      {\hbox{$\scriptscriptstyle\BMfont#1$}}
                 \else{$\BMfont#1$}\fi}

\def\inbar{\vrule height1.5ex width.4pt depth0pt}
\def\sinbar{\vrule height1ex width.35pt depth0pt}
\def\ssinbar{\vrule height.7ex width.3pt depth0pt}
\font\cmss=cmss10
\font\cmsss=cmss10 at 7pt
\def\ZZ{{}Z {\hskip -6.7pt} Z{}}
\def\Ik{\rlx{\rm I\kern-.18em k}}  
\def\IC{\rlx\leavevmode
             \ifmmode\mathchoice
                    {\hbox{\kern.33em\inbar\kern-.3em{\rm C}}}
                    {\hbox{\kern.33em\inbar\kern-.3em{\rm C}}}
                    {\hbox{\kern.28em\sinbar\kern-.25em{\rm C}}}
                    {\hbox{\kern.25em\ssinbar\kern-.22em{\rm C}}}
             \else{\hbox{\kern.3em\inbar\kern-.3em{\rm C}}}\fi}
\def\IP{\rlx{\rm I\kern-.18em P}}
\def\IR{\rlx{\rm I\kern-.18em R}}
\def\IN{\rlx{\rm I\kern-.20em N}}
\def\Ione{\rlx{\rm 1\kern-2.7pt l}}
\def\bbbzz{{\Bbb Z}}

%
\def\unredoffs{} \def\redoffs{\voffset=-.31truein\hoffset=-.59truein}
\def\speclscape{\special{ps: landscape}}

\newbox\leftpage \newdimen\fullhsize \newdimen\hstitle\newdimen\hsbody
\tolerance=1000\hfuzz=2pt\def\fontflag{cm}
\catcode`\@=11 
\hsbody=\hsize \hstitle=\hsize 

\def\nolabels{\def\wrlabeL##1{}\def\eqlabeL##1{}\def\reflabeL##1{}}
\def\writelabels{\def\wrlabeL##1{\leavevmode\vadjust{\rlap{\smash%
{\line{{\escapechar=` \hfill\rlap{\sevenrm\hskip.03in\string##1}}}}}}}%
\def\eqlabeL##1{{\escapechar-1\rlap{\sevenrm\hskip.05in\string##1}}}%
\def\reflabeL##1{\noexpand\llap{\noexpand\sevenrm\string\string%
\string##1}}}
\nolabels
%
\global\newcount\secno \global\secno=0
\global\newcount\meqno \global\meqno=1
\def\newsec#1{\global\advance\secno by1\message{(\the\secno. #1)}
\global\subsecno=0\eqnres@t\noindent{\bf\the\secno. #1}
\writetoca{{\secsym} {#1}}\par\nobreak\medskip\nobreak}
\def\eqnres@t{\xdef\secsym{\the\secno.}\global\meqno=1
\bigbreak\bigskip}
\def\sequentialequations{\def\eqnres@t{\bigbreak}}\xdef\secsym{}
\global\newcount\subsecno \global\subsecno=0
\def\subsec#1{\global\advance\subsecno by1%
\message{(\secsym\the\subsecno.%
 #1)}
\ifnum\lastpenalty>9000\else\bigbreak\fi
\noindent{\it\secsym\the\subsecno. #1}\writetoca{\string\quad
{\secsym\the\subsecno.} {#1}}\par\nobreak\medskip\nobreak}
\def\appendix#1#2{\global\meqno=1\global\subsecno=0%
\xdef\secsym{\hbox{#1.}}
\bigbreak\bigskip\noindent{\bf Appendix #1. #2}\message{(#1. #2)}
\writetoca{Appendix {#1.} {#2}}\par\nobreak\medskip\nobreak}
\def\eqnn#1{\xdef #1{(\secsym\the\meqno)}\writedef{#1\leftbracket#1}%
\global\advance\meqno by1\wrlabeL#1}
\def\eqna#1{\xdef #1##1{\hbox{$(\secsym\the\meqno##1)$}}
\writedef{#1\numbersign1\leftbracket#1{\numbersign1}}%
\global\advance\meqno by1\wrlabeL{#1$\{\}$}}
\def\eqn#1#2{\xdef #1{(\secsym\the\meqno)}\writedef{#1\leftbracket#1}%
\global\advance\meqno by1$$#2\eqno#1\eqlabeL#1$$}
%
\newskip\footskip\footskip8pt plus 1pt minus 1pt
\font\smallcmr=cmr5
\def\footnotefont{\smallcmr}
\def\f@t#1{\footnotefont #1\@foot}
\def\f@@t{\baselineskip\footskip\bgroup\footnotefont\aftergroup%
\@foot\let\next}
\setbox\strutbox=\hbox{\vrule height9.5pt depth4.5pt width0pt} %
\global\newcount\ftno \global\ftno=0
\def\foot{\global\advance\ftno by1\footnote{$^{\the\ftno}$}}
%
\newwrite\ftfile
\def\footend{\def\foot{\global\advance\ftno by1\chardef\wfile=\ftfile
$^{\the\ftno}$\ifnum\ftno=1\immediate\openout\ftfile=foots.tmp\fi%
\immediate\write\ftfile{\noexpand\smallskip%
\noexpand\item{f\the\ftno:\ }\pctsign}\findarg}%
\def\footatend{\vfill\eject\immediate\closeout\ftfile{\parindent=20pt
\centerline{\bf Footnotes}\nobreak\bigskip\input foots.tmp }}}
\def\footatend{}
\global\newcount\refno \global\refno=1
\newwrite\rfile
%
\def\ref{[\the\refno]\nref}%
\def\nref#1{\xdef#1{[\the\refno]}\writedef{#1\leftbracket#1}%
\ifnum\refno=1\immediate\openout\rfile=refs.tmp\fi%
\global\advance\refno by1\chardef\wfile=\rfile\immediate%
\write\rfile{\noexpand\Item{#1}\reflabeL{#1\hskip.31in}\pctsign}%
\findarg\hskip10.0pt}%
\def\findarg#1#{\begingroup\obeylines\newlinechar=`\^^M\pass@rg}
{\obeylines\gdef\pass@rg#1{\writ@line\relax #1^^M\hbox{}^^M}%
\gdef\writ@line#1^^M{\expandafter\toks0\expandafter{\striprel@x #1}%
\edef\next{\the\toks0}\ifx\next\em@rk\let\next=\endgroup%
\else\ifx\next\empty%
\else\immediate\write\wfile{\the\toks0}%
\fi\let\next=\writ@line\fi\next\relax}}
\def\striprel@x#1{} \def\em@rk{\hbox{}}
\def\lref{\begingroup\obeylines\lr@f}
\def\lr@f#1#2{\gdef#1{\ref#1{#2}}\endgroup\unskip}
\def\semi{;\hfil\break}
\def\addref#1{\immediate\write\rfile{\noexpand\item{}#1}} 
%
\def\listrefs{\footatend\vfill\supereject\immediate\closeout%
\rfile\writestoppt
\baselineskip=14pt\centerline{{\bf References}}%
\bigskip{\frenchspacing%
\parindent=20pt\escapechar=` \input refs.tmp%
\vfill\eject}\nonfrenchspacing}
%
\def\listrefsr{\immediate\closeout\rfile\writestoppt
\baselineskip=14pt\centerline{{\bf References}}%
\bigskip{\frenchspacing%
\parindent=20pt\escapechar=` \input refs.tmp\vfill\eject}%
\nonfrenchspacing}
\def\listrefsrsmall{\immediate\closeout\rfile\writestoppt
\baselineskip=11pt\centerline{{\bf References}}
\font\smallreffonts=cmr9 \font\it=cmti9 \font\bf=cmbx9%
\bigskip{ {\smallreffonts%
\parindent=15pt\escapechar=` \input refs.tmp\vfill\eject}}}
\def\listrefsrmed{\immediate\closeout\rfile\writestoppt
\baselineskip=12.5pt\centerline{{\bf References}}
\font\smallreffonts=cmr10 \font\it=cmti10 \font\bf=cmbx10%
\bigskip{ {\smallreffonts%
\parindent=18pt\escapechar=` \input refs.tmp\vfill\eject}}}
\def\startrefs#1{\immediate\openout\rfile=refs.tmp\refno=#1}
\def\xref{\expandafter\xr@f}\def\xr@f[#1]{#1}
\def\refs#1{\count255=1[\r@fs #1{\hbox{}}]}
\def\r@fs#1{\ifx\und@fined#1\message{reflabel %
\string#1 is undefined.}%
\nref#1{need to supply reference \string#1.}\fi%
\vphantom{\hphantom{#1}}\edef\next{#1}\ifx\next\em@rk\def\next{}%
\else\ifx\next#1\ifodd\count255\relax\xref#1\count255=0\fi%
\else#1\count255=1\fi\let\next=\r@fs\fi\next}
\def\figures{\centerline{{\bf Figure Captions}}%
\medskip\parindent=40pt%
\def\fig##1##2{\medskip\item{Fig.~##1.  }##2}}
%

\newwrite\ffile\global\newcount\figno \global\figno=1
\doit0{
\def\fig{fig.~\the\figno\nfig}
\def\nfig#1{\xdef#1{fig.~\the\figno}%
\writedef{#1\leftbracket fig.\noexpand~\the\figno}%
\ifnum\figno=1\immediate\openout\ffile=figs.tmp%
\fi\chardef\wfile=\ffile%
\immediate\write\ffile{\noexpand\medskip\noexpand%
\item{Fig.\ \the\figno. }
\reflabeL{#1\hskip.55in}\pctsign}\global\advance\figno by1\findarg}
\def\listfigs{\vfill\eject\immediate\closeout\ffile{\parindent40pt
\baselineskip14pt\centerline{{\bf Figure Captions}}\nobreak\medskip
\escapechar=` \input figs.tmp\vfill\eject}}
\def\xfig{\expandafter\xf@g}\def\xf@g fig.\penalty\@M\ {}
\def\figs#1{figs.~\f@gs #1{\hbox{}}}
\def\f@gs#1{\edef\next{#1}\ifx\next\em@rk\def\next{}\else
\ifx\next#1\xfig #1\else#1\fi\let\next=\f@gs\fi\next}
}

\newwrite\lfile
{\escapechar-1\xdef\pctsign{\string\%}\xdef\leftbracket{\string\{}
\xdef\rightbracket{\string\}}\xdef\numbersign{\string\#}}
\def\writedefs{\immediate\openout\lfile=labeldefs.tmp %
\def\writedef##1{%
\immediate\write\lfile{\string\def\string##1\rightbracket}}}
\def\writestop{\def\writestoppt%
{\immediate\write\lfile{\string\pageno%
\the\pageno\string\startrefs\leftbracket\the\refno\rightbracket%
\string\def\string\secsym\leftbracket\secsym\rightbracket%
\string\secno\the\secno\string\meqno\the\meqno}%
\immediate\closeout\lfile}}
\def\writestoppt{}\def\writedef#1{}
\def\seclab#1{\xdef #1{\the\secno}\writedef{#1\leftbracket#1}%
\wrlabeL{#1=#1}}
\def\subseclab#1{\xdef #1{\secsym\the\subsecno}%
\writedef{#1\leftbracket#1}\wrlabeL{#1=#1}}
\newwrite\tfile \def\writetoca#1{}
\def\leaderfill{\leaders\hbox to 1em{\hss.\hss}\hfill}
\def\writetoc{\immediate\openout\tfile=toc.tmp
   \def\writetoca##1{{\edef\next{\write\tfile{\noindent ##1
   \string\leaderfill {\noexpand\number\pageno} \par}}\next}}}
\def\listtoc{\centerline{\bf Contents}\nobreak%
 \medskip{\baselineskip=12pt
 \parskip=0pt\catcode`\@=11 \input toc.tex \catcode`\@=12 %
 \bigbreak\bigskip}}
\catcode`\@=12 
%

\countdef\pageno=0 \pageno=1
\newtoks\headline \headline={\hfil}
\newtoks\footline
 \footline={\bigskip\hss\tenrm\folio\hss}
\def\folio{\ifnum\pageno<0 \romannumeral-\pageno \else\number\pageno
 \fi}

\def\nopagenumbers{\footline={\hfil}}
\def\advancepageno{\ifnum\pageno<0 \global\advance\pageno by -1
 \else\global\advance\pageno by 1 \fi}
\newif\ifraggedbottom

\def\raggedbottom{\topskip10pt plus60pt \raggedbottomtrue}
\def\normalbottom{\topskip10pt \raggedbottomfalse}

\def\on#1#2{{\buildrel{\mkern2.5mu#1\mkern-2.5mu}\over{#2}}}
\def\dt#1{\on{\hbox{\bf .}}{#1}}                
\def\Dot#1{\dt{#1}}

\def\eqdot{~{\buildrel{\hbox{\LARGE .}} \over =}~}
\def\eqstar{~{\buildrel * \over =}~}
\def\eqques{~{\buildrel ? \over =}~}

\def\lhs{({\rm LHS})}
\def\rhs{({\rm RHS})}
\def\lhsof#1{({\rm LHS~of~({#1})})}
\def\rhsof#1{({\rm RHS~of~({#1})})}

\def\binomial#1#2{\left(\,{\buildrel
{\raise4pt\hbox{$\displaystyle{#1}$}}\over
{\raise-6pt\hbox{$\displaystyle{#2}$}}}\,\right)}

\def\Dsl{{}D \!\!\!\! /{\,}}
\def\doubletilde#1{{}{\buildrel{\mkern1mu_\approx\mkern-1mu}%
\over{#1}}{}}

\def\hata{{\hat a}} \def\hatb{{\hat b}}
\def\hatc{{\hat c}} \def\hatd{{\hat d}}
\def\hate{{\hat e}} \def\hatf{{\hat f}}

\def\circnum#1{{\ooalign%
{\hfil\raise-.12ex\hbox{#1}\hfil\crcr\mathhexbox20D}}}

\def\Christoffel#1#2#3{\Big\{ {\raise-2pt\hbox{${\scst #1}$}
\atop{\raise4pt\hbox{${\scst#2~ #3}$} }} \Big\} }



\font\smallcmr=cmr6 scaled \magstep2
\font\smallsmallcmr=cmr5 scaled \magstep 1
\font\largetitle=cmr17 scaled \magstep1
\font\LargeLarge=cmr17 scaled \magstep5
\font\largelarge=cmr12 scaled \magstep0

\def\alephnull{\aleph_0}
\def\sqrtoneovertwopi{\frac1{\sqrt{2\pi}}\,}
\def\twopi{2\pi}
\def\sqrttwopi{\sqrt{\twopi}}

\def\rmA{{\rm A}} \def\rmB{{\rm B}} \def\rmC{{\rm C}}
\def\HatC{\Hat C}

\def\alpr{\a{\hskip 1.2pt}'}
\def\dim#1{\hbox{dim}\,{#1}}
\def\leftarrowoverdel{{\buildrel\leftarrow\over\partial}}
\def\rightarrowoverdel{{\buildrel\rightarrow\over%
\partial}}
\def\ee{{\hskip 0.6pt}e{\hskip 0.6pt}}

\def\neq{\not=}
\def\lowlow#1{\hskip0.01in{\raise -7pt%
\hbox{${\hskip1.0pt} \!_{#1}$}}}

\def\atmp#1#2#3{Adv.~Theor.~Math.~Phys.~{\bf{#1}}
(19{#2}) {#3}}

\font\smallcmr=cmr6 scaled \magstep2

\def\fracmm#1#2{{{#1}\over{#2}}}
\def\fracms#1#2{{{\small{#1}}\over{\small{#2}}}}
\def\low#1{{\raise -3pt\hbox{${\hskip 1.0pt}\!_{#1}$}}}

\def\ip{{=\!\!\! \mid}}
\def\Lslash{${\rm L}{\!\!\!\! /}\, $}

\def\leapprox{~\raise 3pt \hbox{$<$} \hskip-9pt \raise -3pt \hbox{$\sim$}~}
\def\geapprox{~\raise 3pt \hbox{$>$} \hskip-9pt \raise -3pt \hbox{$\sim$}~}

\def\vev#1{\langle{#1} \rangle}
\def\sigmaslashI{\S_I \hskip -11pt \raise1pt\hbox{/}{} \,\,\,}
\def\SigmaslashI{\sum_I \!\!\!\!\! / \,}
\def\sigmaslashi{\S_i \hskip -10pt \raise1pt\hbox{/}{} \,\,}
\def\Sigmaslashi{\sum_i \!\!\!\!\! / \,\,}

\mathchardef\undertilde="0366
\def\underTilde#1{\!\raise -10pt\hbox{$\undertilde{~}$}\hskip-14pt{#1}{}}

\def\framing#1{\doit{#1}  {\framingfonts{#1}
\border\headpic  }}

\framing{0}



\doit0{
{\bf Preliminary Version (FOR YOUR EYES
ONLY!)\hfill\today
} \\[-0.25in]
\\[-0.3in]
}

\doit0{
{\hbox to\hsize{\hfill
hep-th/yymmnnn}}
\vskip -0.06in
}

\doit1{
{\hbox to\hsize{\hfill CSULB--PA--07--4}}
\vskip -0.14in
}

\hfill
\\

\vskip 0.8in

\begin{center}

{\large\bf Comment on Electroweak Higgs as a} \\
{\large\bf Pseudo-Goldstone Boson of Broken Scale Invariance}  \\

\baselineskip 9pt

\vskip 0.5in

Hitoshi ~N{\smallcmr ISHINO}%
\footnotes{E-Mail: hnishino@csulb.edu} and
~Subhash ~R{\smallcmr AJPOOT}%
\footnotes{E-Mail: rajpoot@csulb.edu}
\\[.16in]  {\it Department of Physics \& Astronomy}
\\ [.015in]
{\it California State University} \\ [.015in]
{\it 1250 Bellflower Boulevard} \\ [.015in]
{\it Long Beach, CA 90840} \\ [0.02in]

\vskip 3.8in

{\bf Abstract}\\[.1in]
\end{center}

\vskip 0.1in

\baselineskip 14pt

The first model of Foot, Kobakhidze and Volkas  described in their work in
arXiv:0704.1165 [hep-ph] is  a tailored version of our model on
broken scale invariance in the standard model presented in
\hepth{0403039}.


\baselineskip 8pt

\doit0{
\leftline{\small PACS: }
\vskip 0.06in
\leftline{\small Key Words:   }
\leftline{\small {\hskip 0.8in} ?????}
}

\vfill\eject

\oddsidemargin=0.03in
\evensidemargin=0.01in
\hsize=6.5in
\topskip 0.46in
\textwidth=6.5in
\textheight=9in
\flushbottom
\footnotesep=1.0em
\footskip=0.32in
\def\baselinestretch{0.8}

\baselineskip 20pt

\pageno=2



The merits of implementing scale invariance in the standard model of
particle interactions were enunciated  by us in
\ref\nroriginal{H.~Nishino and S.~Rajpoot, {\it `Broken Scale Invariance in the Standard Model'}, CSULB-PA-04-2, \hepth{0403039}.}.
An extended version of this work that also addresses the important
issue of unification of elementary particle interactions will appear
in
\ref\nrrecent{H.~Nishino and S.~Rajpoot, {\it `Standard Model and
SU(5) GUT with Local Scale Invariance and the Weylon'},
CSULB-PA-06-4,
to appear in {\it CICHEP-II Conference Proceedings}, 2006, published by AIP.}.
The  salient features  of our model were  recently recapitulated in
our comment
\ref\nrcomment{H.~Nishino and S.~Rajpoot,
{\it `Comment on Shadow and Non-Shadow Extensions of the Standard Model'},
\hepth{0702080}.}.

Here we  point out that in the  work of Foot, Kobakhidze and Volkas
\ref\fkv{R.~Foot, A.~Kobakhidze and R.R.~Volkas,
{\it `Electroweak Higgs as a Pseudo-Goldstone Boson
of Broken Scale Invariance'}, arXiv:0704.1165 [hep-ph].}
on broken scale invariance in the standard model, their first model
corresponds to a tailored version of our model.

R.~Foot et.~al.~\fkv\ are fully aware of the fact that  scale
invariance symmetry can be realised as a local
symmetry,{\footnotes{Cf.~Footnote \#2 in \fkv.}} in which
case breaking it results in an additional neutral gauge boson.  In
our work \nroriginal\ we dubbed this gauge boson the Weylon, named
after Herman Weyl
\ref\weyl{H.~Weyl, S.-B.~Preuss. Akad.~Wiss.~{\textbf {465}} (1918);
Math.~Z.~{\textbf 2} (1918) 384; Ann.~Phys.~{\textbf {59}} (1919) 101;
{\it Raum, Zeit, Materie, vierte erweiterte Auflage}:
Julius Springer (1921).}.

\vskip 2.3in


\immediate\closeout\rfile\writestoppt
\baselineskip=14pt\centerline{{\bf References}}%
\bigskip{\frenchspacing%
\parindent=20pt\escapechar=` \input refs.tmp\vfill\eject}%
\nonfrenchspacing

\end{document}